\documentclass[aps,pre,amsmath,amssymb,floatfix]{revtex4}

\usepackage{graphicx}
\usepackage{epstopdf}
\usepackage{pstricks}
\usepackage{xcolor}
\usepackage{caption}
\usepackage{subcaption}
\usepackage{amsmath}
\usepackage{amssymb}
\usepackage{mathrsfs}
\usepackage{verbatim}

\newcommand{\iu}{\mathrm{i}} % imaginary unit

\usepackage[figcolor=white]{todonotes}

\newcommand{\TJambul}[1]{\todo[inline, color=green!60]{Jambul: #1}}
\newcommand{\TMatthias}[1]{\todo[inline, color=blue!40]{Matthias: #1}}

\begin{document}

\title{Transparent boundary conditions for the spatially discrete Schr\"odinger equation: Reflectionless quantum transport in 1D lattices}
% \title{Exact Transparent Boundary Conditions for the Spatially Discrete Schrödinger Equation on 1D Lattices}
% \title{Reflectionless Quantum Transport in 1D Lattices via Discrete Transparent Boundary Conditions}
% \title{Analytical Derivation and Numerical Discretization of Transparent Boundary Conditions for the Discrete 1D Schrödinger Equation}

\author{M.E.~Akramov$^{1,*}$, J.R.~Yusupov$^{2}$, M.~Ehrhardt$^{3}$, and D.U.~Matrasulov$^{4}$}
\affiliation{$^1$National University of Uzbekistan, Universitet Str. 4, 100174, Tashkent, Uzbekistan\\
$^2$Kimyo International University in Tashkent, 156 Usman Nasyr Str., 100121, Tashkent, Uzbekistan\\
$^3$Bergische Universit\"at Wuppertal, Gau{\ss}strasse 20, D-42119 Wuppertal, Germany\\
$^4$Turin Polytechnic University in Tashkent, 17
Niyazov Str., 100095, Tashkent, Uzbekistan}
\email[Corresponding author: ]{mashrabresearcher@gmail.com}
%%%%%%%%%%%%%%%%%%
\begin{abstract}
We construct exact transparent boundary conditions (TBCs) for a time-continuous, spatially discrete Schrödinger equation that models a one-dimensional quantum lattice. 
Using a recently developed exact solution for the discrete system, we derive the Dirichlet-to-Neumann maps analytically via Laplace transforms.
This yields a convolution-type boundary condition governed by Bessel functions.
We rigorously demonstrate the consistency of this discrete formulation with its continuous counterpart in the continuum limit. 
Additionally, we present an efficient time-discretization scheme based on the trapezoidal rule for practical implementation. 
Numerical experiments using a Crank-Nicolson solver verify that our proposed TBCs eliminate spurious backscattering entirely and preserve reflectionless propagation of a Gaussian wave packet exiting the computational domain.
\end{abstract}

\maketitle

%%%%%%%%%%%%%%%%%%%%%%%%%%%%% Introduction %%%%%%%%%%%%%%%%%%%%%%%%
\section{Introduction}
Low-dimensional discrete structures have attracted significant attention in the context of functional quantum materials and devices. 
A critical challenge in this field is realizing tunable quantum transport, 
which is directly relevant to designing advanced optoelectronic materials and optimizing solid-state quantum devices. To address this challenge, we develop physically realistic models that allow for precise control over quasiparticle dynamics in the quantum regime. 

The main goal of this control is to eliminate backscattering during particle transport to prevent signal degradation and energy loss. In quantum mechanics, the conventional approach to minimizing backscattering is to construct a scattering matrix whose reflection coefficients vanish. However, this approach is less effective when managing transmission directly across artificial boundaries and provides fewer real-time control tools. Consequently, robust alternative methods must be developed to ensure reflectionless propagation at specific domain boundaries. One highly successful approach is the concept of ``absorbing boundary conditions", also referred to as ``artificial" or ``transparent" boundary conditions (TBCs), which were pioneered in \cite{Engquist1, Engquist2} and has since been expanded upon in a broad body of literature \cite{Halpern, Sofronov, Schmidt1997, Arnold1998, Ehrhardt1999, Ehrhardt2001, Arnold2003, Antoine2008, Ehrhardt3, Antoine09, Antoine10, Ehrhardt2025}. 

The core principle of TBCs is based on deriving specific boundary constraints for a given evolution equation. These constraints allow a wave or particle to propagate within a truncated computational domain as if it were moving through infinite space. In essence, the boundary is mathematically "invisible," enabling seamless, reflectionless transmission. Despite their mathematical rigor and accuracy, analytical TBCs often have highly complex, nonlocal forms that complicate their implementation in physical systems. 
In certain contexts, however, it is possible to uncover simple, physically intuitive conditions that reduce the TBC to natural boundary conditions. 
These reductions have been demonstrated for various evolution equations on metric graphs \cite{Jambul1, Jambul2, Aripov, Jambul4, Jambul3, TBCSGE, mashrab, Jambul5, mashrab2}, 
where the constraints simplify into an elegant sum rule matching continuity and Kirchhoff conditions at a graph vertex.

Recently, these approaches were extended to stationary Schrödinger equations by leveraging quantum graphs \cite{Jambul6} and Weyl-Titchmarsh theory \cite{Derkach2025}. 
However, despite the vast spectrum of applications for the TBC framework, almost all existing literature is restricted to continuous evolution equations, and discrete lattice equations have largely remained outside the scope of such exact formulations. 
In this paper, we address this issue by developing the TBC framework for the time-dependent, spatially discrete linear Schrödinger equation. 
Building upon an exact analytical solution rather than a numerical discretization of a continuous boundary allows us to provide a mathematically exact and computationally robust framework for reflectionless lattice transport. 

The paper is organized as follows. In the next section we will present the concept 
of transparent boundary conditions both for continuous and discrete Schrodinger equations. The section includes also derivation of the continuous limit and discretization of TBC. Numerical implementation of TBC is presented in section III. Section IV provides some concluding remarks.

%%%%%%%%%%%%%%%%%%%%%%%%%%%%%
\section{The Concept of Transparent Boundary Conditions}
Transparent boundary conditions (TBCs) were first developed by Engquist and Majda for the acoustic equation \cite{Engquist1, Engquist2}. 
Since then, the TBC framework has been successfully extended to other parabolic and hyperbolic partial differential equations (PDEs) \cite{Halpern, Sofronov}. 
The concept has been thoroughly examined in a series of studies across various mathematical and physical contexts: \cite{Schmidt1997, Arnold1998, Ehrhardt1999, Ehrhardt2001, Arnold2003, Antoine2008, Ehrhardt3, Antoine09, Antoine10, Klein2010, Klein2011, KleinThesis}. 
For the linear Schrödinger equation, Shibata and Kuska formulated approximate TBCs using plane-wave dispersion relations to truncate the computational domain \cite{Shibata,Kuska}.  

However, most existing studies focus on discretizing continuous TBCs alongside the governing differential equation rather than deriving them directly from exact solutions. 
Recently, an exact solution was proposed for the discrete Schrödinger equation on a one-dimensional lattice \cite{DSE2024}. 
This solution provides explicit analytical expressions for the wave function, enabling the direct investigation of spectral and transport properties without introducing approximations associated with numerical discretization. 
The existence of this exact solution implies that boundary conditions can be derived directly from the underlying discrete equation rather than through a finite-difference approximation of a continuous boundary.
This motivates developing transparent boundary conditions based on exact analytical representations, which are expected to remain transparent regardless of the numerical scheme chosen for subsequent time computations.

%%%%%%%%%%%%%%%%%%%%%%%%%%%%%
\subsection{Transparent Boundary Conditions for the 1D Schr\"odinger Equation: Continuous Case}\label{sec:continuous}
Before introducing the discrete model, it is instructive to recall the standard formulation of TBCs on a line for the continuous case.
Consider the motion of a quantum wave (or particle) in a 1D domain $x \in \mathbb{R}$, governed by the time-dependent Schrödinger equation (setting $\hbar = m = 1$): 
\begin{equation}\label{lse03}
   \iu\frac{\partial}{\partial t} \Psi=-\frac12\frac{\partial^2}{\partial x^2} \Psi
   + V(x,t)\Psi,\quad x\in\mathbb{R}, \;t>0,
\end{equation}  
subject to the initial condition
%\begin{equation*}
    $\Psi(x,0)=\Psi^I(x)$,
%\end{equation*}
   %with the initial data 
   where $\Psi^I\in L^2(\mathbb{R})$ and an external potential $V(.,t)\in L^\infty(\mathbb{R})$.
   % and $V(x,t)$ is an external potential.

The objective is to establish boundary conditions (BCs) for \eqref{lse03} at the endpoints $x = 0$ and $x = L$ that allow waves to pass out of the computational domain without reflection.
If the solution restricted to the interval $[0, L]$ under these artificial boundary conditions coincides exactly with the unrestricted solution across the whole space, the boundary conditions are classified as transparent.

The systematic procedure for designing these conditions was established in Refs.~\cite{Ehrhardt1999, Ehrhardt2001, Arnold2003, Antoine2008} and is summarized below:
\begin{enumerate}
\item Split the original problem into coupled equations: interior and exterior problems. 
\item Apply a Laplace transformation in time $t$.
\item Solve the ordinary differential equations in $x$.
\item Allow only “outgoing” waves by selecting the decaying solution as $x\to\pm\infty$. 
\item Match the Dirichlet and Neumann values at $x=0$, $x=L$.
\item Apply the inverse Laplace transformation.
\end{enumerate} 

One can apply this procedure to derive the right TBC at $x=L$ using the following two basic assumptions:
The initial data, denoted by $\Psi^I$, is compactly supported in the computational domain, $0<x<L$.
The given external potential is constant outside this finite domain,
i.e.\ $V(x,t)=0$ for $x\leqslant 0$, 
$V(x,t)=V_L$ for $x\geqslant L$, 
one can derive the right TBC at $x=L$:
\begin{equation}\label{tbcconR}
  \frac{\partial}{\partial x}\Psi(L,t)
  = -\sqrt{\frac{2}{\pi}}\mathrm{e}^{-\iu\frac{\pi}{4}}\mathrm{e}^{-\iu V_Lt}\frac{d}{dt}\int\limits_0^t{\frac{\Psi(L,\tau)\mathrm{e}^{\iu V_L\tau}}{\sqrt{t-\tau}}\,d\tau}.
\end{equation}
Similarly, the left TBC at $x=0$ is obtained as
\begin{equation}\label{tbcconL}
  \frac{\partial}{\partial x}\Psi(0,t)
  = \sqrt{\frac{2}{\pi}} \mathrm{e}^{-\iu\frac{\pi}{4}} \frac{d}{dt} \int\limits_0^t{\frac{\Psi(0,\tau)}{\sqrt{t-\tau}}\,d\tau}.
\end{equation}
For a comprehensive review of the derivation of these continuous TBCs \eqref{tbcconR} and \eqref{tbcconL}, see \cite{Antoine2008}.

%%%%%%%%%%%%%%%%%%%%%%%%%%%%%
\subsection{Transparent Boundary Conditions for Spatially Discrete Schr\"odinger Equation}
The discrete Schr\"odinger equation describes motion of quantum particle in 1D lattice.
We will focus on the time-continuous %dependent 
spatially discrete Schr\"odinger equation for the wave function $\Psi_j=\Psi_j(t)$:
\begin{equation}\label{sdse}
    \iu\frac{d \Psi_j}{dt} + \frac{1}{2h^2} \bigl(\Psi_{j-1}-2\Psi_j+\Psi_{j+1}\bigr) = 0,  \quad j\in\mathbb{Z},\;t>0, 
\end{equation}
considered in the infinite 1D lattice ($-\infty<j<\infty$) with uniform grid size $h=\Delta x$. 
The initial condition is given as
%\begin{equation*}
    $\Psi_j(0) = \Psi_j^I$,
%\end{equation*}
where $\Psi_j^I\in \ell^2(\mathbb{Z})$. 
The semi-discrete equation \eqref{sdse} can be regarded as the result of a spatial standard finite difference discretization of the (free) linear Schr\"odinger equation, or a discrete quantum
system with an infinite number of degrees of freedom.
Here, our motivation is to consider \eqref{sdse} as a model appearing in charge carriers dynamics in  conducting polymers and molecular chains, where electron-hole bound system moves along the spatially discrete structure cf.\ 
\cite{Hegger2020,Exciton2020,Chernyak1,Chernyak2,Chernyak3}.

% We want to 
We will construct the boundary conditions for Eq.~\eqref{sdse} that provide reflectionless transmission of the wave at the boundaries of the finite discrete interval, $0\leq j\leq J$. 
To do this, we use the same procedure described in % the previous
Subsection~\ref{sec:continuous} for the continuous case.
Thus, we split the whole discrete domain $x_j=jh$, $j\in\mathbb{Z}$, into ``interior" $0\leq j\leq J$, and two ``exterior" problems determined by $j\leq 0$ and $j\geq J$ \cite{Antoine2008}.
The ``interior" problem reads with $h=L/J$: 
\begin{equation}\label{interior}
\begin{split}
    \iu\frac{d \Psi_j}{dt} &+ \frac{1}{2h^2} \bigl(\Psi_{j-1}-2\Psi_j+\Psi_{j+1}\bigr) = 0, \quad j=1,\dots,J-1, \; t>0, \\
     \Psi_j(0) &= \Psi_j^I, \\
     D_x^+ \Psi_0(t) &:= \frac{1}{h} \bigl[ \Psi_1(t)-\Psi_0(t) \bigr] = (T_0 \Psi_0)(t), \\
     D_x^- \Psi_J(t) &:= \frac{1}{h} \bigl[ \Psi_J(t)-\Psi_{J-1}(t) \bigr] = (T_J \Psi_J)(t),
\end{split}
\end{equation}
where $T_0$ and $T_J$ are called Dirichlet-to-Neumann maps at the boundaries of the discrete interval $x_j\in [0, L]$, $L=Jh$.

\begin{figure}[htb]
    \centering
    \includegraphics[width=0.4\linewidth]{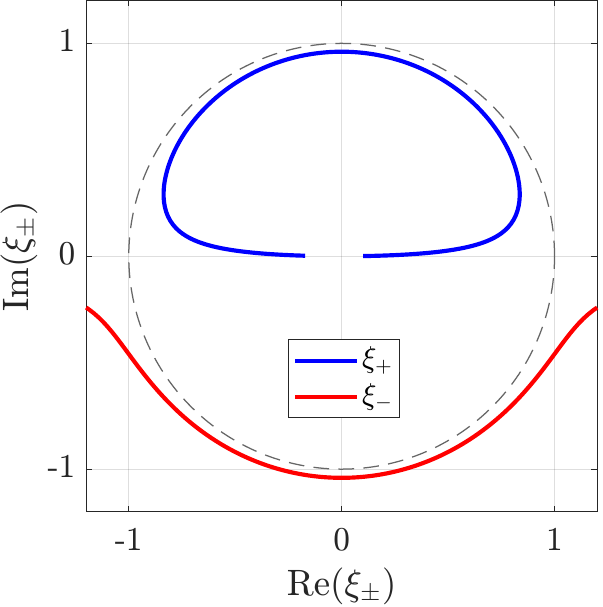}
    \caption{Trajectories of $\xi_{\pm}(s)$ in the complex plane for $h=0.2$. 
    The blue curve ($\xi_+$) lies inside the unit circle, while the red curve ($\xi_-$) lies outside, confirming that the plus sign in Eq.~\eqref{exterior_char_sol} is the physically admissible root satisfying 
    % with 
    the decay condition $|\xi_+| \le 1$.}
    \label{fig:xi}
\end{figure}

The right ``exterior" problem for $j\geq J$ can be written as
\begin{equation} \label{exterior}
\begin{split} 
    \iu\frac{d \Phi_j}{dt} &+ \frac{1}{2h^2} \bigl( \Phi_{j-1}-2\Phi_j+\Phi_{j+1} \bigr) = 0, \quad j>J, \; t>0, \\
    \Phi_J(0) &= 0, \\
    \Phi_J(t) &= \Psi_J(t), \quad \lim_{j\to\infty}\Phi_j(t) = 0, \\
    D_x^+ \Phi_J(t) 
      &:= \frac{1}{h} \bigl[ \Phi_{J+1}(t)-\Phi_J(t) \bigr]=(T_J\Phi_J)(t),
\end{split}
\end{equation}
with $\Phi(t)\in\ell^2(J,\infty)$, for all $t>0$. %, similar for $\Psi$
The current conservation condition on the boundaries imply
\begin{equation}
    D_x^- \Psi_J(t) = D_x^+ \Phi_J(t), \quad
    D_x^+ \Psi_0(t) = D_x^- \Phi_0(t).    
\end{equation}
Next, we apply the Laplace transformation in time
\begin{equation}
    \hat{\Phi}_j(s):=\mathcal{L}(\Phi_j)(s) 
    := \int_0^{\infty} \Phi_j(t) \,\mathrm{e}^{-st} \,dt
\end{equation} 
to the right exterior problem \eqref{exterior}, which yields 
\begin{equation} \label{exterior_laplace}
\begin{split} 
    &\frac{1}{2h^2} \bigl(\hat{\Phi}_{j-1}(s) -2\hat{\Phi}_j(s) +\hat{\Phi}_{j+1}(s)\bigr) + \iu s \hat{\Phi}_j(s) = 0, \quad j>J, \\
    %%  \Phi_J(0) &= 0,  yields no extra term in the Laplace transform
    &\hat{\Phi}_J(s) = \hat{\Psi}_J(s), \quad \lim_{j\to\infty}\hat{\Phi}_j(s) = 0.
    % & (T_J\Phi_J)(t) &= D_x \Phi_J(t) = \frac{1}{a} \bigl[ \Phi_{J+1}(t)-\Phi_J(t) \bigr].
\end{split}
\end{equation}
The transformed exterior problem \eqref{exterior_laplace} is a homogeneous second order difference equation with constant coefficients. 
% Thus, the solutions have a power form 
Inspired by the 
exact power-form solution of the discrete Schr\"odinger equation derived 
in \cite{DSE2024}, we seek the solution of the form
$\hat{\Phi}_j(s)=\xi(s)^{j-J}$, where $\xi(s)$ solves the characteristic equation
\begin{equation} \label{exterior_char}
    \xi^2(s) -2(1-\iu h^2s)\,\xi(s) +1 = 0, \quad j>J, 
\end{equation} 
i.e.\ the two fundamental solutions $\xi_{\pm}(s)$ are given by
  \begin{equation} \label{exterior_char_sol}
    \xi_{\pm}(s)= 1-\iu h^2s \pm \iu h\sqrt{2\iu s +h^2s^2}.  
\end{equation}   
Since the solution must decay as $j \to \infty$ and $\xi_+(s) \cdot \xi_-(s) = 1$ from Vieta's formulas, exactly one root satisfies $|\xi(s)| \leq 1$. As shown in Fig.~\ref{fig:xi}, this root is $\xi_+(s)$, and we therefore select the plus sign in Eq.~\eqref{exterior_char_sol}. 
This leads to the transformed right transparent boundary condition (TBC)
\begin{equation}
    \Hat{\Phi}_{J+1}(s) = \xi_+(s) \,\Hat{\Phi}_J(s), 
\end{equation}
i.e. the Laplace-transformed Dirichlet-to-Neumann operator
$T_J$ reads
\begin{equation}\label{eq1}
     D_x^+ \hat{\Phi}_J(s) 
    := \frac{1}{h}  \Bigl[ \xi_+(s)-1 \Bigr] \hat{\Phi}_J(s) 
    = \Bigl[ -\iu h s + \iu\sqrt{2\iu s+h^2s^2} \Bigr] \hat{\Phi}_J(s)
    =:(T_J\hat{\Phi}_J)(s).
\end{equation}

Now, we compute the inverse Laplace transform of $D_x^+ \hat{\Phi}_J(s)$ using three tools: the Bessel function Laplace pair, the first shifting, and the convolution rules.
From standard tables, the following Laplace pair holds for $a > 0$:
\begin{equation}\label{bessel}
    \mathcal{L}\left\{\frac{J_1(at)}{t}\right\}(s) 
    = \frac{-s + \sqrt{s^2 + a^2}}{a}.
\end{equation}
We rewrite the square root as
\begin{equation*}
  \sqrt{-2\iu s - h^2s^2} = \iu h\sqrt{\Bigl(s + \frac{\iu}{h^2}\Bigr)^2 + \frac{1}{h^4}},
\end{equation*}
and using \eqref{bessel} with phase shifting rule gives
\begin{equation*}
\sqrt{-2\iu s - h^2s^2} = \iu hs - \frac{1}{h} + \frac{\iu}{h}\,\mathcal{L}\Bigl\{\mathrm{e}^{-\frac{\iu}{h^2}t}\frac{J_1(t/h^2)}{t}\Bigr\}(s).
\end{equation*}
The convolution rule
\begin{equation*}
    \mathcal{L}^{-1}\{\hat{F}(s)\cdot\hat{\Phi}_J(s)\} 
    = \int_0^t F(t-\tau)\,\Phi_J(\tau)\,d\tau,
\end{equation*}
yields to obtain the final expression of the inverse Laplace transform given by
\begin{equation}\label{tbc1}
    D_x^- \Psi_J(t) = - \frac{1}{h} \Psi_J(t) 
    % - \iu h\,\frac{d}{dt} \Psi_J(t) 
    +\frac{\iu}{h} \int_0^t \mathrm{e}^{-\iu\frac{t-\tau}{h^2}}  \frac{1}{t-\tau}J_1\Bigl(\frac{t-\tau}{h^2}\Bigr)  \Psi_J(\tau) \,d\tau,
\end{equation}
with the Bessel function of the first kind and order 1.

\subsection{Consistency with Continuous Counterpart}
One may also be interested in verifying the consistency of the obtained result \eqref{tbc1} with its continuous counterpart \eqref{tbcconR}. This can be achieved by considering the continuum limit, i.e., by taking the limit as $h\to 0$. 
To this end, we present a detailed derivation of 
\begin{equation}\label{eq:lim}
\lim_{h\to 0}D_x^- \Psi_J(t) 
     = \lim_{h\to 0}\Big[- \frac{1}{h} \Psi_J(t)  
    +\frac{\iu}{h} \int_0^t \mathrm{e}^{-\iu\frac{t-\tau}{h^2}}  \frac{1}{t-\tau}J_1\Bigl(\frac{t-\tau}{h^2}\Bigr)  \Psi_J(\tau) \,d\tau\Big].
\end{equation}
Denoting the convolution integral on the r.h.s.\ of \eqref{eq:lim} as $I$, we rewrite
\begin{equation}\label{limint}
\frac{\partial}{\partial x}\Psi(x_J,t) = \lim_{h\to 0}\Big[-\frac{1}{h} \Psi_J(t)  
    +\frac{\iu}{h} I\Big].
\end{equation}

Next, introducing the new variable $u=(t-\tau)/h^2$, the integral becomes
\begin{equation}
   I=\int_0^{t/h^2} \frac{\mathrm{e}^{-\iu u}}{u} J_1(u)\Psi_J(t-h^2u)\,du.
\end{equation}
For the cases $h=0$ and $h>0$ with $h\to0$, the integral can be rewritten as
\begin{equation}\label{I2}
   I= \Psi_J(t)\int_0^{\infty} \frac{\mathrm{e}^{-\iu u}}{u} J_1(u)\,du
     +\int_0^t \frac{\mathrm{e}^{-\iu \frac{v}{h^2}}}{v} 
   J_1\Bigl(\frac{v}{h^2}\Bigr)\Psi_J(t-v)\,dv,
\end{equation}
where $v=h^2u$.

Now, the first integral is known (from the standard table of Laplace transforms):
\begin{equation}
    \int_0^{\infty} \frac{\mathrm{e}^{-s u}}{u} J_1(u)\,du=\sqrt{s^2+1}-s.
\end{equation}
In our case $s=\iu$, which gives the value $-\iu$.
For large arguments, the Bessel function can be approximated by (see \cite[Formula~9.2.1]{AbrSt65})
\begin{equation}
  J_1\Bigl(\frac{v}{h^2}\Bigr)
  \approx h\sqrt{\frac{2}{\pi v}}\cos\Bigl(\frac{v}{h^2}-\frac{3\pi}{4}\Bigr).
\end{equation}

Substituting this approximation into \eqref{I2} and neglecting the rapidly oscillating term, we obtain
\begin{equation}
I = -\iu \Psi_J(t) +\frac{h}{2}\sqrt{\frac{2}{\pi}}\mathrm{e}^{-\iu \frac{3\pi}{4}}
      \int_0^t \frac{\Psi_J(t-v)}{v^{3/2}}\,dv.
\end{equation}

Thus, Eq.~\eqref{limint} reads
\begin{equation}\label{limint1}
\frac{\partial}{\partial x}\Psi(x_J,t) = \frac{1}{2}\sqrt{\frac{2}{\pi}}
\mathrm{e}^{-\iu \frac{3\pi}{4}}
\int_0^t \frac{\Psi_J(t-v)}{v^{3/2}}dv.
\end{equation}

Introducing the change of variables
$\tau=t-v$,
%$v=t-\tau$,
%$dv=-d\tau$,
we obtain
\begin{equation}
\int_0^t \frac{\Psi_J(t-v)}{v^{3/2}}dv
=-2\int_0^t\Psi_J(\tau)\frac{d}{d t}(t-\tau)^{-1/2}\,d\tau.
\end{equation}

Using the relationship between Riemann-Liouville and Caputo fractional derivatives 
\cite{Podlubny,Huseynov},
% \cite{Podlubny},
for the $1/2$-derivative, one has
\begin{equation*}
\frac{1}{\Gamma(1/2)}\int_0^t\Psi_J(\tau)\frac{d}{d t}(t-\tau)^{-1/2}\,d\tau=\frac{1}{\Gamma(1/2)}\dfrac{d}{dt}\int_0^t \dfrac{\Psi_J(\tau)}{\sqrt{t-\tau}}\,d\tau - \frac{\Psi_J(0)}{\Gamma(1/2)\sqrt{t}}.
\end{equation*}
As it was assumed that the initial data is compactly supported in the computational domain ($0<j<J$), the value of the wave function at $t=0$ on the boundary vanishes, i.e., $\Psi_J(0)=0$. Thus, in our case, the Caputo and Riemann-Liouville fractional derivatives coincide, and one obtains 
% of order $1/2$ \cite{Samko1993, Podlubny1999}:
\begin{equation}
    \int_0^t \frac{\Psi_J(t-v)}{v^{3/2}}\,dv = 
    -2\,\dfrac{d}{dt}\int_0^t \dfrac{\Psi_J(\tau)}{\sqrt{t-\tau}}\,d\tau.
\end{equation}
% Using the Riemann--Liouville fractional derivative, 
%the integral can be rewritten as
%\begin{equation}
%I = -\iu \Psi_J(t)
%+\iu h\sqrt{\frac{2}{\pi}}
%\mathrm{e}^{-\iu \frac{\pi}{4}}
%\frac{d}{dt}
%\int_0^t\frac{\Psi_J(\tau)}{\sqrt{t-\tau}}\,d\tau.
%\end{equation}
%
Substituting this result into the TBC in Eq.~\eqref{tbc1}, 
we obtain the continuous TBC for the Schr\"odinger equation~\eqref{tbcconR}:
\begin{equation}
  \frac{\partial}{\partial x}\Psi(x_J,t)
   =-\sqrt{\frac{2}{\pi}}\mathrm{e}^{-\iu \frac{\pi}{4}}
   \frac{d}{dt} \int_0^t \frac{\Psi(x_J,\tau)}{\sqrt{t-\tau}}\,d\tau.
\end{equation}

%%%%%%%%%%%%%%%%%%%%%%%%%%%%%%%%%%%%%%%
\subsection{Discretization of the TBC}
The backward finite-difference operator at the boundary node is defined by
\begin{equation}
D_x^- \Psi_J(t)
=\frac{1}{h}\bigl(\Psi_J(t)-\Psi_{J-1}(t)\bigr).
\end{equation}
Using this operator, the TBC can be written in the form
%\begin{equation}
   $2\Psi_{J}(t)-\Psi_{J-1}(t)=\mathcal{I}(t)$,
%\end{equation}
where
\begin{equation}
    \mathcal{I}(t)=\frac{\iu}{h}\int_0^tK(t-\tau)\Psi_J(\tau)\,d\tau,\quad
    \text{with}\quad
    K(t)=\mathrm{e}^{-\iu\,t/h^2}\frac{1}{t}J_1\Bigl(\frac{t}{h^2}\Bigr)
\end{equation}
is the convolution term and $K(t)$ denotes the corresponding kernel function.
%\begin{equation}
%    K(t)=\mathrm{e}^{-\iu\,t/h^2}\frac{1}{t}J_1\Bigl(\frac{t}{h^2}\Bigr)
%\end{equation}
%denotes the corresponding kernel function.
%
To discretize the convolution integral in time, the trapezoidal rule is employed:
\begin{equation}
   \int_0^t f(t)\,dt
   \approx\frac{\Delta t}{2}\bigl(f_0+f_N\bigr)+\sum_{p=1}^{N-1}f_p\,\Delta t.
\end{equation}
Applying this quadrature formula to the convolution term at the time level $t_n$ yields
\begin{equation}
\mathcal{I}_n=\frac{\iu}{h}\frac{\Delta t}{2}
\Bigl[K(t_n)\Psi_{0,J}+K(0)\Psi_{n,J}\Bigr]
+\frac{\iu}{h}\Delta t\sum_{p=1}^{n-1}K(t_n-\tau_p)\Psi_{p,J}.
\end{equation}
Substituting the above expression into the transparent boundary condition gives
\begin{equation}
   2\Psi_{n,J}-\Psi_{n,J-1}=\iu\frac{\Delta t}{2}
  \Bigl[K(t_n)\Psi_{0,J}+K(0)\Psi_{n,J}\Bigr]
  +\iu\Delta t\sum_{p=1}^{n-1}K(t_n-\tau_p)\Psi_{p,J}.
\end{equation}
% Collecting all terms containing $\Psi_{n,J}$ on the left-hand side leads to
% \begin{equation}
% \Psi_{n,J}
% \left(
% 1+\frac{\Delta t^2}{4h^2}K(0)
% \right)
% =
% \Psi_{n-1,J}
% -
% \frac{i\Delta t}{2h^2}\Psi_{n,J-1}
% -
% \frac{\Delta t^2}{4h^2}
% K(t_n)\Psi_{0,J}
% -
% \frac{\Delta t^2}{2h^2}
% \sum_{p=1}^{n-1}
% K(t_n-\tau_p)\Psi_{p,J}.
% \end{equation}
Collecting all terms containing $\Psi_{n,J}$ yields the explicit formula
\begin{equation}\label{tbc_discrete}
   \Psi_{n,J}=\frac{1}{2-\iu\frac{\Delta t}{4h^2}}
   \biggl[\Psi_{n,J-1}+\iu\frac{\Delta t}{2}K(t_n)\Psi_{0,J}
     +\iu\Delta t\sum_{p=1}^{n-1}K(t_n-\tau_p)\Psi_{p,J}\biggr].
\end{equation}
Here, using the small-argument asymptotic of the Bessel function $J_{\nu}(z)\sim \frac{1}{\Gamma(\nu+1)} \bigl(\frac{z}{2}\bigr)^{\nu}$ as $z \to 0$ (see Formula~9.1.7 of Ref.~\cite{AbrSt65}), we can define the limiting value of the kernel at the origin, which is given by
% \textcolor{red}{reference?}
\begin{equation}
    K(0)=\lim_{t\to0} \mathrm{e}^{-\iu t/h^2}\frac{J_1(t/h^2)}{t}=\frac{1}{2h^2}.
\end{equation}

%%%%%%%%%%%%%%%%%%%%%%%%%%%%%%%%%%%%%%%% figure
\begin{figure}[t]
    \centering
    \includegraphics[width=0.6\textwidth]{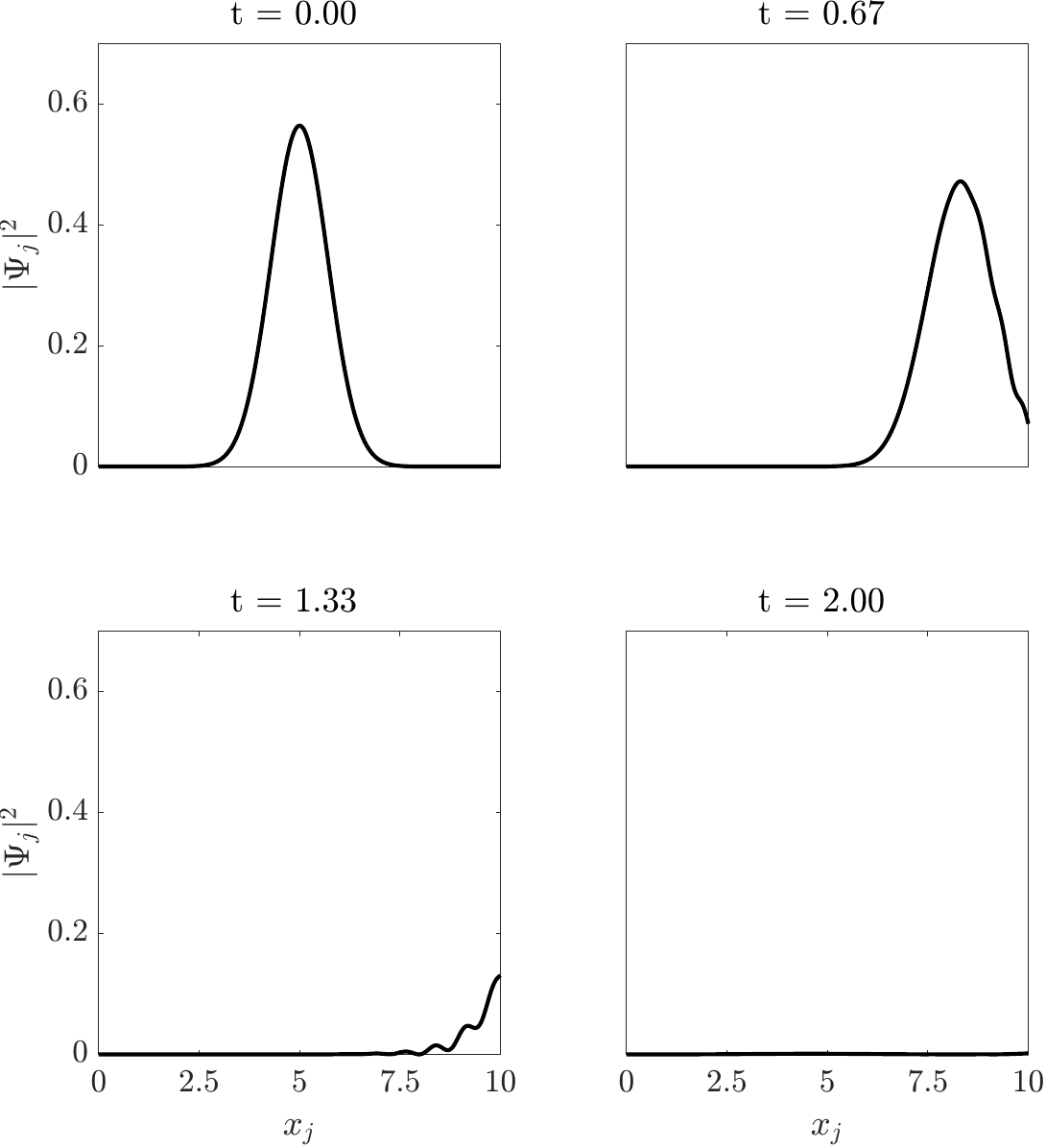}
    \caption{Snapshots of $|\Psi_j(t)|^2$ at $t = 0,\, 0.67,\, 1.33,\, 2.00$. 
    The Gaussian wave packet propagates to the right and leaves the computational domain through the transparent boundary without being reflected back.}
    \label{fig:profile}
\end{figure}

%%%%%%%%%%%%%%%%%%%%%%%%%%%%%%%%%%%%%% Section 3 %%%%%%%%%%%%%%%%
\section{Numerical Experiment}
To verify the accuracy and mathematical fidelity of the derived TBCs, we numericall solve the spatially discrete Schrödinger equation~\eqref{sdse} on the discrete finite interval $x_j \in [0, L]$ with $L = 10$. 
For the temporal evolution, we use the second-order accurate Crank-Nicolson scheme. For the interior nodes ($j = 1, \dots, J-1$), the resulting finite-difference scheme is expressed as:  
\begin{equation}
\frac{\Psi_j^{n+1} - \Psi_j^n}{\Delta t} 
= \frac{\iu}{4h^2}\bigl(\Psi_{j-1}^{n+1} - 2\Psi_j^{n+1} + \Psi_{j+1}^{n+1}
+\Psi_{j-1}^{n} - 2\Psi_j^{n} + \Psi_{j+1}^{n}\bigr),
\end{equation}
where $\Psi_j^n \approx \Psi_j(t_n)$ denotes the numerical approximation of the wave function at the discrete time level $t_n = n\Delta t$. 
At each time step, the resulting tridiagonal linear system is efficiently solved using the Thomas algorithm.
The right boundary value $\Psi_J^{n+1}$ is updated explicitly using the time-discretized TBC \eqref{tbc_discrete}. The left boundary condition is set to a homogeneous Dirichlet condition, $\Psi_0^n = 0$, for all $n$.  
%
\begin{comment}
where $\Psi_j^n \approx \Psi_j(t_n)$ and $t_n = n\Delta t$. 
At each time step the resulting tridiagonal linear system is solved. 
% efficiently by LU decomposition. 
The right boundary value $\Psi_J^n$ is updated explicitly using the discretized TBC given 
in Eq.~\eqref{tbc_discrete}.
% where the kernel $K(t)$ is defined in Eq.~\eqref{kernel} and 
% evaluated using the Bessel function $J_1$. 
The left boundary condition is set to $\Psi_0^n = 0$ for all $n$.
\end{comment}
%
The initial condition is chosen to be a Gaussian wave packet centered at the midpoint of the computational domain ($x_0 = L/2$):
\begin{equation}
     \Psi_j^0 = A\exp\Bigl(-\frac{(x_j - x_0)^2}{2\sigma^2}\Bigr)\,\mathrm{e}^{\iu k_0 x_j},
\end{equation}
with a width of $\sigma = 1$, a wave number of $k_0 = 5$, and $A$ a normalization constant.
The discretization parameters are $J = 400$, $h = 0.025$, and a time step $\Delta t = 6.25 \cdot 10^{-6}$.  

To quantitatively evaluate the performance and transparency of the TBC, we monitor the time evolution of the discrete norm: % (total probability):  
\begin{equation}\label{norm}
    M(t) = h\sum_{j=0}^{J}|\Psi_j(t)|^2.
\end{equation}
For an exact whole-space solution, this norm remains identically equal to one ($M(0)=1$). As the wave packet exits the finite computational domain through the transparent boundaries, $M(t)$ should decay monotonically to zero without experiencing any artificial backward reflections.

\begin{figure}[t!]
    \centering
    \includegraphics[width=0.35\textwidth]{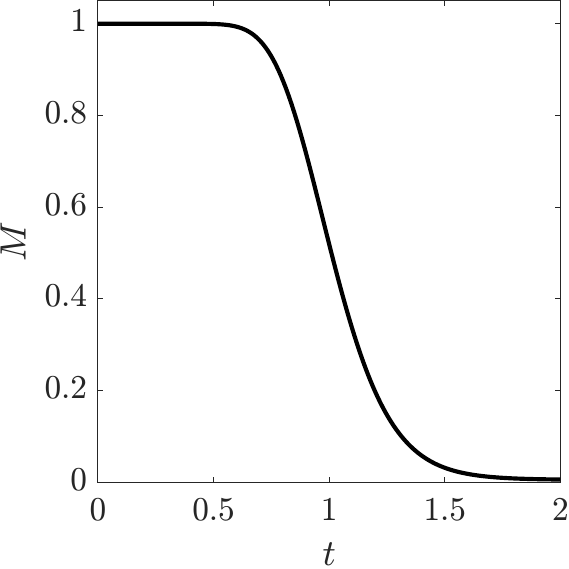}
    \caption{Time evolution of the discrete norm in Eq.~\eqref{norm}.}
    \label{fig:norm}
\end{figure}

The numerical results are illustrated in Figs.~\ref{fig:profile} and \ref{fig:norm}.
Figure~\ref{fig:profile} displays snapshots of the probability density $|\Psi_{j}(t)|^{2}$ at four representative times ($t = 0.00$, $0.67$, $1.33$, and $2.00$).
The wave packet stably propagates to the right, smoothly leaving the computational domain through the right boundary with no visible backscattering or grid-induced reflections. 
This confirms the reflectionless nature of the derived discrete TBC.  

This observation is further supported by Fig.~\ref{fig:norm}, which presents the temporal evolution of the discrete norm $M(t)$. 
As the wave packet exits the computational domain, the norm smoothly and monotonically decreases from unity to zero. There is no oscillatory behavior or late-time increase in the norm, demonstrating that the artificial boundary introduces no significant numerical reflections.

\section{Conclusions}
In this work, we performed a comprehensive analysis of the time-dependent, spatially discrete Schrödinger equation within the TBC framework. 
Using an exact analytical solution recently developed for 1D lattices, we derived explicit, non-local Dirichlet-to-Neumann maps in the time domain. 
Unlike standard approaches, which introduce grid-dependent artifacts when discretizing continuous boundary conditions, our formulation is derived directly at the discrete level.

We demonstrated that in the continuum limit, as the lattice grid spacing approaches zero, our discrete TBCs perfectly recover the well-known fractional-derivative boundary conditions of the continuous Schrödinger equation. 
To facilitate practical implementation, we presented an efficient time discretization scheme that uses a trapezoidal quadrature rule for the history-dependent convolution term. 
High-resolution numerical simulations using a Crank-Nicolson solver confirmed the accuracy of our approach. 
The simulations showed that a Gaussian wave packet injected into a finite computational domain exits smoothly without generating unphysical backscattering or numerical reflections and preserves the discrete norm.

The exact "transparent" quantum lattice model established in this paper provides a robust theoretical framework for analyzing open quantum systems without artificial boundaries.
Looking forward, this discrete boundary formalism can be directly applied to model unhindered, high-fidelity quasiparticle and exciton transport in realistic, low-dimensional structures. 
Such structures include branched conducting polymers, topological insulators, and engineered quantum wires, where backscattering suppression is essential for optimizing devices. 
Future work will focus on extending this exact discrete TBC methodology to network geometries, such as discrete quantum graphs, as well as exploring the impact of localized nonlinearities.

%%%%%%%%%%%%%%%%%%%%%%%%%%%%%%%%%%%%%%%% figure

%%%%%%%%%%%%%%%%%%%%%%%%%%%%%
\section*{Acknowledgements}
We acknowledge the funding provided by
the Grant of the Innovations Agency under the Ministry of Higher Education, Research and Innovations (FL-8824063336). 
The work of JY and DM is also supported by a grant from the Innovation Development Agency of the Republic of Uzbekistan (Ref.\ No.\ F-2021-440).

%%%%%%%%%%%%%%%%%%%%%%%%%%%%% Refs

\end{document}